\newcommand{\GG}[1]{\textcolor{black}{#1}}
\newcommand{\G}[1]{\textcolor{black}{#1}}

\documentclass[]{aa}
\usepackage{graphicx,url,twoopt,natbib}
\usepackage[varg]{txfonts}
\usepackage{subfigure} 
\usepackage{enumerate}

\bibpunct{(}{)}{;}{a}{}{,} 
\usepackage[colorlinks,
            linkcolor=red,
            anchorcolor=blue,
            citecolor=blue]{hyperref}
\graphicspath{{./}{figures/}}

\begin{document}

\title{Non-local thermodynamic equilibrium (NLTE) abundances of europium (Eu) for a sample of metal-poor stars in the galactic halo and metal-poor disk with 1D and $<$3D$>$ models}
\author{{Yanjun Guo}\inst{1,2,3},
          Nicholas Storm\inst{2,4},
          Maria Bergemann\inst{2},
          Jianhui Lian\inst{2,5},
          Sofya Alexeeva\inst{6},
          Yangyang Li\inst{7},
          Rana Ezzeddine\inst{7,8},
          Jeffrey Gerber\inst{2},
          \and
          XueFei Chen\inst{1}
          }
\institute{
           $^1$ Yunnan observatories, Chinese Academy of Sciences, P.O. Box 110, Kunming, 650011, China; guoyanjun@ynao.ac.cn\\
            $^2$ Max-Planck Institute for Astronomy, Königstuhl 17, D-69117 Heidelberg, Germany\\
            $^3$ International Centre of Supernovae, Yunnan Key Laboratory, Kunming 650216, China\\
            $^4$ Heidelberg University, Grabengasse 1, 69117 Heidelberg, Germany;\\
            $^5$ South-Western Institute for Astronomy Research, Yunnan University, Kunming, Yunnan 650091, People’s Republic of China\\
            $^6$ CAS Key Laboratory of Optical Astronomy, National Astronomical Observatories, Chinese Academy of Sciences, Beijing, 100101, People’s Republic of China\\  
            $^7$ Department of Astronomy, University of Florida, Bryant Space Science Center, Gainesville, FL 32611, USA\\   
            $^8$ Joint Institute for Nuclear Astrophysics - Center for Evolution of the Elements, USA\\
             }

\abstract
   {As a key to chemical evolutionary studies, the distribution of elements in galactic provides a wealth of information to understand the individual star formation histories of galaxies.
The r-process is a complex nucleosynthesis process, and the origin of r-process elements is heavily debated.
Europium (Eu) is viewed as an almost pure r-process element.
Accurate measurements of europium abundances in cool stars are essential for an enhanced understanding of the r-process mechanisms.
}
   {We measure the abundance of Eu in solar spectra and a sample of metal-poor stars in the Galactic halo and metal-poor disk, with the metallicities ranging from \GG{$-2.4$} to $-0.5$ dex, using non-local thermodynamic equilibrium (NLTE) line formation. We compare these measurements with Galactic Chemical Evolution (GCE) models to \GG{explore the impact of the NLTE corrections on the contribution of r-process site in Galactic chemical evolution.} }
    {In this work, we used NLTE line formation, as well as one-dimensional (1D) hydrostatic and spatial averages of three-dimensional hydrodynamical ($<$3D$>$) model atmospheres to measure the abundance of Eu based on both the Eu II 4129 \AA\ and Eu II 6645 \AA\ lines for solar spectra and metal-poor stars.
    }
    {We find that \GG{for Eu II 4129 \AA\ line the NLTE modeling leads to higher (0.04 dex) solar Eu abundance in 1D and higher (0.07 dex) in \GG{$<$3D$>$} NLTE while} NLTE modeling leads to higher (0.01 dex) solar Eu abundance in 1D and lower (0.03 dex) in \GG{$<$3D$>$} NLTE for Eu II 6645 \AA\ line. 
    \GG{Although the NLTE corrections for the Eu II $\lambda$ 4129 \AA\ and Eu II $\lambda$ 6645 \AA\ lines are opposite, the discrepancy between the abundances derived from these individual lines reduces after applying NLTE corrections, highlighting the critical role of NLTE abundance determinations.}
    By comparing these measurements with Galactic chemical evolution (GCE) models, we find that the \G{amount of NLTE correction does not require significant change of the parameters for Eu production} in GCE models.} 
    {}


\titlerunning{NLTE abundances of Eu} 
\authorrunning{Yanjun Guo et al.}        
\maketitle

\section{Introduction} \label{sec:intro}
Abundance ratios as a function of time or metallicity  provide important information about the scenarios of nucleosynthesis and also allow us to trace the chronology of events in Galaxy, since different chemical elements are produced on different time scales in different astrophysical sites \citep{1991Rana,1994Wilson,2003Chiappini,2012Matteucci,2014Matteucci}.
Iron peak elements such as iron, cobalt, and chromium, are produced over longer time-scales in Type Ia supernovae \citep{1957Burbidge,1994Nomoto,2000Hillebrandt,2013Seitenzahl,2019Bergemann}. 
In addition, the abundance of $\alpha$-elements such as magnesium and oxygen has been demonstrated to be a sensitive \GG{proxy for the initial phases} of \G{Galactic chemical evolution}, which are dominated by the production of heavy elements and the return from core-collapse supernovae \citep{1980Tinsley,2015Woosley}. 
Slow neutron-capture process (s-process) elements, such as strontium and barium, are produced in intermediate and low-mass asymptotic giant branch (AGB) stars \citep{1980Tinsley,1997Pagel,2001Matteucci,2014karakas,2016Frischknecht,2018choplin}.
Similar, rapid neutron-capture process (r-process) elements, such as europium, provide insight into violent events such as neutron star mergers, neutron-driven winds in core-collapse SNe, explosions of rapidly rotating
magnetised massive stars  \citep{1999Rosswog,1994Takahashi,1994Woosley,arcones2013,bliss2018a,siegel2019,2023Reichert}.

While numerous studies have provided detailed insights into the chemical evolution of $\alpha$-elements and iron peak elements, there is a paucity of theoretical and observational data available for neutron capture elements, especially for the r-process elements \citep{2015Woosley,2016Frischknecht,2018choplin,2023Lian}.
The r-process is a complex process that requires extremely high neutron density, making the origin of r-process elements remains a topic of ongoing debate in the scientific community \citep{1957Burbidge,martin2015,siegel2017,cote2018a,halevi2018,radice2018,siegel2019}.
Europium is almost a pure r-process element \citep{2008Sneden,2021Cowan}. 
Thus, precise measurements of Eu abundances in cool stars are crucial for constraining the sites of nucleosynthetic production.

\cite{2018Guiglion} investigated three pure r-process elements, including Eu, Gd, and Dy based on MARCS model atmosphere and the local thermodynamic equilibrium (LTE) code.
They found that the [Eu/Fe] ratio follows a continuous sequence from the thin disk to the thick disk with respect to metallicity. 
\citet{2023Lian} compared the evolution of the [Eu/Fe]-[Fe/H] trend, based on Eu abundance measurements using 1D LTE model, with the GCE model in the metal-rich region. 
However, their study lacks a metal-poor sample and relies solely on 1D LTE results.
As shown in recent literature studies \citep{2014Bergemann,2014Bergemann24element,2017BergemannMg,2023Storm,2024A&AStorm}, it is crucial to take into account the effects of NLTE and 3D effects for FGKM-type stars.
Therefore, further studies based on the application of NLTE and $<$3D$>$ corrections to Eu abundances in a metal-poor sample is essential.

The first analysis of NLTE effects in Eu can be traced back to \cite{2000Mashonkina}, which presents the statistical equilibrium of Eu II using a model atom that includes 32 levels of Eu II, along with the ground state of Eu III. This approach derived the solar europium abundance with A(Eu) = 0.53 dex.
\cite{2016zhaogang} followed the atomic model from \cite{2000Mashonkina} and investigated Eu abundances in 1D NLTE analysis.
Recently, \citet{2024A&AStorm} published the latest Eu atom model with 163 levels of Eu II to investigate the solar abundance of Eu.
In this work, we apply NLTE line formation, as well as 1D hydrostatic and $<$3D$>$ model atmospheres with the latest Eu atomic data from \citet{2024A&AStorm} to measure the abundance of Eu of \GG{164} metal-poor stars in the Galactic halo and metal-poor disk with [Fe/H] ranging from \GG{$-2.4$} to $-0.5$ dex adopted from \cite{2011ApJRuchti}. 

The structure of the paper is as follows. 
We introduce our observed data in Section.~\ref{sec:data}. 
In Section \ref{sec:method}, we describe the detail of method we use to analysis the abundance of Eu.
The best-fit results for the Sun and the comparison with GCE model are presented in Sec.~\ref{sec:floats}.
Finally, we summarize our conclusions in Sec.~\ref{sec:Summary}.

\section{Data} \label{sec:data}
The sample is adopted from \cite{2011ApJRuchti}. 
It includes 319 metal-poor stars in Galactic disk and halo with the effective temperature ($T_\mathrm{eff}$) ranging from $4050$ to $6500$ K, the surface gravity ($\log{g}$) ranging from $0.5$ to $4.5$, the metallicities ranging from $-2.8$ to $-0.5$ dex.
These stars were selected to investigate thick disk-like kinematics, and they were originally observed by the RAVE survey \citep{2006Steinmetz}.
The sample was observed using several high-resolution facilities, including the MIKE spectrograph on the Magellan-Clay telescope \citep{2003Bernstein}, FEROS on the MPG 2.2 m telescope \citep{1999Kaufer}, UCLES spectrograph on the Anglo-Australian telescope \citep{1985Walker}, and the ARC spectrograph on the Apache Point 3.5 m telescope \citep{2003Wang}.
The wavelength range of the spectra observed from UCLES is from 4460 to 7270 \AA, whereas the other three cover the full optical range from 3500 to $\sim$ 9500 \AA.
The signal-to-noise ratios (S/Ns) of all the spectra are greater than 100 pixel$^{-1}$ at around 6000 \AA.
We adopted the NLTE-opt stellar parameters from Table 1 in \cite{2017Bergemann} as the input data for further analysis.

\section{Methods}\label{sec:method}
Here, we provide a description of the model atmospheres, spectral synthesis code, NLTE models, and the linelist we use in this work.
\subsection{Model atmospheres}\label{sec:Model Atmospheres}
We used two \G{grids of} models to analyze our metal-poor sample: the 1D line-blanketed hydrostatic MARCS model from \cite{2008Gustafsson} and the average 3D model (hereafter, $<$3D$>$) Stagger model from \cite{2013Magic1,2013Magic2}.

The MARCS model is a homogeneous model atmosphere for late-type (FGKM) stars, assuming LTE. This model uses the thermal equilibrium and Saha equation for all number densities of atoms and molecules \citep{1928Gibson,1934Russell}, as well as the Boltzmann distribution for all partition functions and excitation equilibrium \citep{1981Irwin,2008Gustafsson}. 
The MARCS model atmospheres provided $\approx16000$ standard composition model atmospheres\footnote{\url{ https://marcs.astro.uu.se}} covering a wide parameter space:
the range of effective temperature ($T_{\rm eff}$) goes from 2500K to 8000K, with a step of 100K for ($T_{\rm eff}$) from 2500K to 4000K and a step of 250K for all others.
For the surface gravity (log g), the range is from $-1$ to $5.0$ dex with a step of $0.5$ dex.
The range of metallicity ([Fe/H]) is from $-5.0$ to $1.0$, with a step of $0.25$ dex from $-1.0$ to $1.0$ dex, $0.5$ dex from $-3.0$ to $1.0$, and $1$ dex from $-5.0$ to $-3.0$.
The micro-turbulence, with four different values ($0$, $1$, $2$, and $5$ km $s^{-1}$) is also included in the grids.

The grid of Stagger model atmospheres is a collection of three-dimensional (3D) and time-dependent hydrodynamic model atmospheres with a more realistic treatment of the radiative transfer equation for late-type stars \citep{2013Magic1,2013Magic2}.
The grid of $<$3D$>$ Stagger model atmospheres used in this work is an average of such sets of Stagger model atmospheres on surfaces of equal optical depth (log$\tau_{5000}$) also adopted from \citep{2013Magic1,2013Magic2}.
The grid models presented by \cite{2013Magic1} consist of approximately 220 models, which can be found on their website\footnote{\url{https://staggergrid.wordpress.com}}. 
The models cover the following ranges of stellar parameters: effective temperature ($T_{\rm eff}$) from 4000 K to 7000 K with a step size of $500$ K, surface gravity (log g) from $1.5$ to $5.0$ dex with a step size of $0.5$ dex, and metallicity ([Fe/H]) from $-4.0$ to $0.5$ dex with a step size of $0.5$ dex from $-1$ to $0.5$ dex, and $1.0$ dex from $-4.0$ to $-1$ dex.

\subsection{Spectral synthesis code}
Turbospectrum (TS) is a spectral synthesis code based on LTE radiative transfer \citep{1964Feautrier,1984Nordlund,1998Alvarez,2012Plez}. 
The TS code has been continuously developed over the years, and recently the latest version v20.0\footnote{\url{https://github.com/bertrandplez/Turbospectrum2020}} was published.
The most significant update in this version is the ability to generate NLTE spectra.
Similar to other abundance analysis and spectrum synthesis codes, this is achieved by using grids of NLTE departure coefficients to calculate the NLTE line profiles by correcting the line source functions and the line opacity of all lines \citep[further details described in][]{2023Gerber}.

A Python wrapper called Turbospectrum Spectral Fitting with Python (TSFitPy)\footnote{\url{https://github.com/JGerbs13/TSFitPy}} developed by \cite{2023Gerber} has specifically been designed to determine stellar abundances, optional with other parameters, such as micro-turbulence ($\xi_{t}$) using the Nelder-Mead (simplex algorithm) minimization method \citep{10.1093/comjnl/7.4.308, 35200b4570854615b7fb6cabbba0f2b6}.
TSFitPy was notably updated to fit either macroturbulence or rotation for each individual line using Limited-memory Broyden-Fletcher-Goldfarb-Shanno (L-BFGS-B) with bound consideration algorithm \citep{doi:10.1137/0916069, 10.1145/279232.279236} using the Scipy Python package \citep{2020SciPy-NMeth}. 
This was done as a secondary step after generating synthetic spectra with a specific stellar abundance to break the degeneracy of fitting the abundance and the broadening of lines simultaneously \citep{2023Storm}.
\GG{A dedicated interpolating function is provided together with TS code, which takes the rectangular grids of model atmospheres and the corresponding grids of departure coefficients, and produces an interpolated atmosphere structure and departure file for a desired combination of stellar parameters \citep{2023Gerber}.}

In this work, we used the updated TSFitPy and TS code to fit the observed spectra and obtain the stellar abundances of the metal-poor sample.

\subsection{NLTE models}\label{sec: NLTE calculate}

\begin{table}
\caption{\label{tab:Eu model}Eu II line used for the abundance calculations.}
\centering
\begin{tabular}{cccc}
 \hline \hline
$\lambda$ &   Lower  &  Upper  & $\rm log \it gf$\\ \AA & level & level  &   \\ \hline
4129.7    & $a^{9}S_{4}^{0}$  &  $z^{9}P_{4}$  & 0.22 \\
6645.1    & $a^{9}D_{6}^{0}$  &  $z^{9}P_{5}$  & 0.12 \\  
\hline 
\end{tabular}\\
\end{table}

\begin{figure}
    \centering
    \includegraphics[scale=0.7]{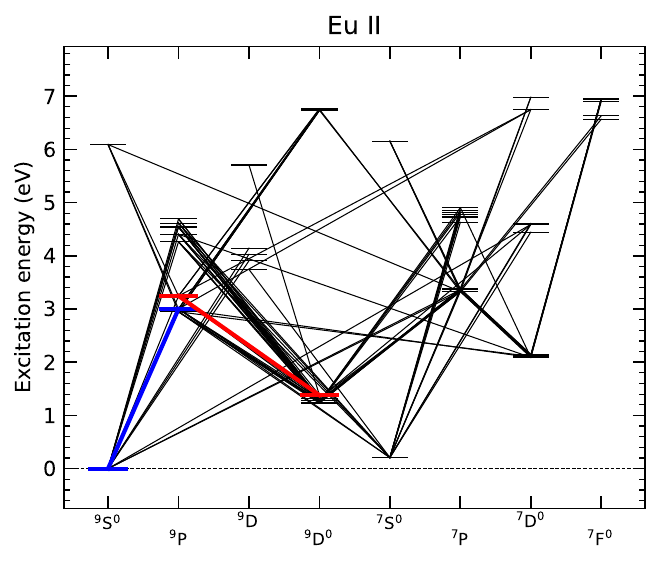}
    \caption{Grotrian diagram of the \GG{Eu II model atom}. The model atom are taken from the \GG{\cite{2024A&AStorm}}. The \GG{blue} and red lines represent the transitions giving rise to the \GG{Eu II 4129 \AA\ } and 6645 \AA\ lines, respectively.}
    \label{fig: Grotrian diagrams}
\end{figure}

We adopted the NLTE model of Eu from \citet{2024A&AStorm}, which utilizes data from the NIST\footnote{\url{https://physics.nist.gov/PhysRefData/ASD/levels_form.html}} and Kurucz\footnote{\url{http://kurucz.harvard.edu/atoms/6301/}} databases \citep{1978Martin,2000Nakhate,2017Johnson}.
The electronic structure of Eu comprises a total of 662 levels, with Eu I having 498 levels and Eu II having 163 levels. 
The ionization potentials for Eu I and Eu II are 5.67 eV and 11.24 eV, respectively \citep{1978Martin}. 
The atom model of Eu contains three ionisation stages and is closed by the Eu III state.
The Grotrian diagram of the \GG{Eu II model atom} is shown in Fig.~\ref{fig: Grotrian diagrams}.

We used the MULTI1D code to compute statistical equilibrium (SE) calculations, which was developed by \cite{1986Carlsson}.
This code solves the the SE equations iteratively and handles the radiative transfer equation in a 1D plane-parallel geometry, assuming that deviations from LTE do not influence the structure of the input model atmosphere, which underlies the standard assumption of a trace element to calculate the statistical equilibrium of NLTE elements.
Recently, MULTI1D was updated by our group \citep{2019Bergemann,2020Gallagher} and widely used for NLTE analyses of atmospheric parameters and chemical abundances \citep{2023Gerber,2023Storm, 2023Li}.

We used a Python wrapper\footnote{\url{https://github.com/stormnick/wrapper_multi}} for MULTI1D to calculate the departure coefficients of Eu for both grids of MARCS and $<$3D$>$ Stagger model atmospheres \G{\citep[using the same methodology as in Sec. 2.4 in ][]{2023Gerber}. }
For each individual star, we obtained the abundance directly using TS by fitting the spectra, where the NLTE synthetic spectra are based on the precomputed departure coefficients from MULTI1D.
\G{As detailed, the europium departure coefficient grids are computed between [Eu/Fe] = $-2$ to +1 in steps of 0.1 dex. 
During the fitting process, the closest departure coefficient within [Eu/Fe] step is used ( within 0.05 dex) for the synthetic spectra generation.
For example a star with [Eu/Fe] = 0.6 dex would use departure coefficients computed with that specific abundance. Therefore, different [Eu/Fe] abundances are taken into account during the statistical equilibrium calculations for each individual star.}

\subsection{Linelist and line choice}\label{sec: Linelist}
We used the homogeneous Gaia-ESO linelist, which contains atomic and molecular data from \cite{2021Heiter} and was recently updated by \cite{2022Magg} with atomic data for several elements.
For the solar Eu abundance, we adopted A(Eu) = 0.57 from \citet{2024A&AStorm}.
For the Eu II lines, the $gf$-values were measured by \citet{2001Lawler} based on experimental life-times and branching fractions (BFs).

\GG{Eu II 4129.73 \AA\ is the resonance line most widely used in Eu abundance determinations \citep{2000Mashonkina,2007Franccois,2016zhaogang,2024Lucchesi}.
Eu II 6645.10 \AA\ is the strongest Eu II line in the yellow-red spectral region and is partially blended with weak Si I lines at 6645.21 \AA\ in the solar spectrum \citep{2001Lawler}. It is also reliably used in stellar Eu abundance studies \citep{2000Mashonkina,2001Lawler,2021Heiter,2024A&AStorm}.}
Therefore, the final abundance analysis primarily relies on these two Eu II line feature. 
The atomic parameters of these two lines are provided in
Table.~\ref{tab:Eu model}.

There are two odd isotopes for Eu, namely, $^{151}\rm Eu$ and $^{153}\rm Eu$.
The solar isotopic abundance ratio for this two isotopes is set to 47.8:52.2, respectively from \cite{2009Lodders}.
\GG{The Eu II 4129\AA\ line is represented by 11 individual hyperfine splitting (HFS) components from \cite{2000Mashonkina}.}
The Eu II 6645\AA\ line is represented by 11 hyperfine and isotopic components \citep{2024A&AStorm}.

\section{Results} \label{sec:floats}     
\subsection{Synthetic spectra}
\begin{figure}
	\centering
	\includegraphics[scale=0.7]{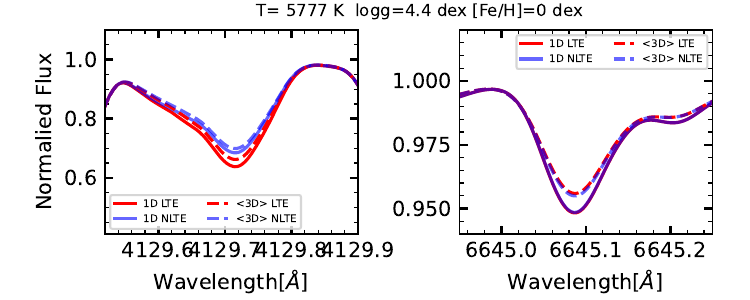}
    \caption{Synthetic spectra for \GG{Eu II 4129\AA\ (left) and} Eu II 6645\AA\ (right) generated from TSFitPy base on solar parameter with A(Eu)=0.57 dex. The red solid lines represent the line profile generated from 1D LTE and the blue solid lines are from 1D NLTE, while the red dash lines represent the line profile generated from $<$3D$>$ LTE and the blue dash lines are from $<$3D$>$ NLTE.}
    \label{fig: synthetic Eu}
\end{figure} 

To showcase the TSFitPy capability, we generated the spectra with 1D LTE/NLTE and $<$3D$>$ LTE/NLTE lines profiles based on solar parameters for \GG{Eu II 4129\AA\ (left) and} Eu II 6645\AA\ (right) in Figure \ref{fig: synthetic Eu}.
The red solid lines represent line profiles generated from 1D LTE, while the blue solid lines are from 1D NLTE. The red dashed lines represent line profiles generated from $<$3D$>$ LTE, and the blue dashed lines are from $<$3D$>$ NLTE. 
\GG{From the comparison of the red and blue lines in the left panel, we observe that the NLTE effect weakens the Eu II 4129\AA\ line in both 1D and $<$3D$>$ models.}
By comparing the red and blue solid lines in the right panel, we find that the NLTE effect slightly weakens the line in the 1D model.
This result is consistent with the positive 1D NLTE correction found by \citet{2000Mashonkina} and \citet{2024A&AStorm}.
In contrast, when comparing the red and blue dashed lines, we observe that the NLTE effect strengthens the line in the $<$3D$>$ model.
\citet{2024A&AStorm} studied the impact of full 3D NLTE modeling of Eu on solar abundances. 
They found that the Eu NLTE model results in a slightly positive correction in 1D and a negative correction in 3D, which is consistent with our findings.

\subsection{Solar Eu abundance}
\begin{figure}
    \centering
    \includegraphics[scale=0.5]{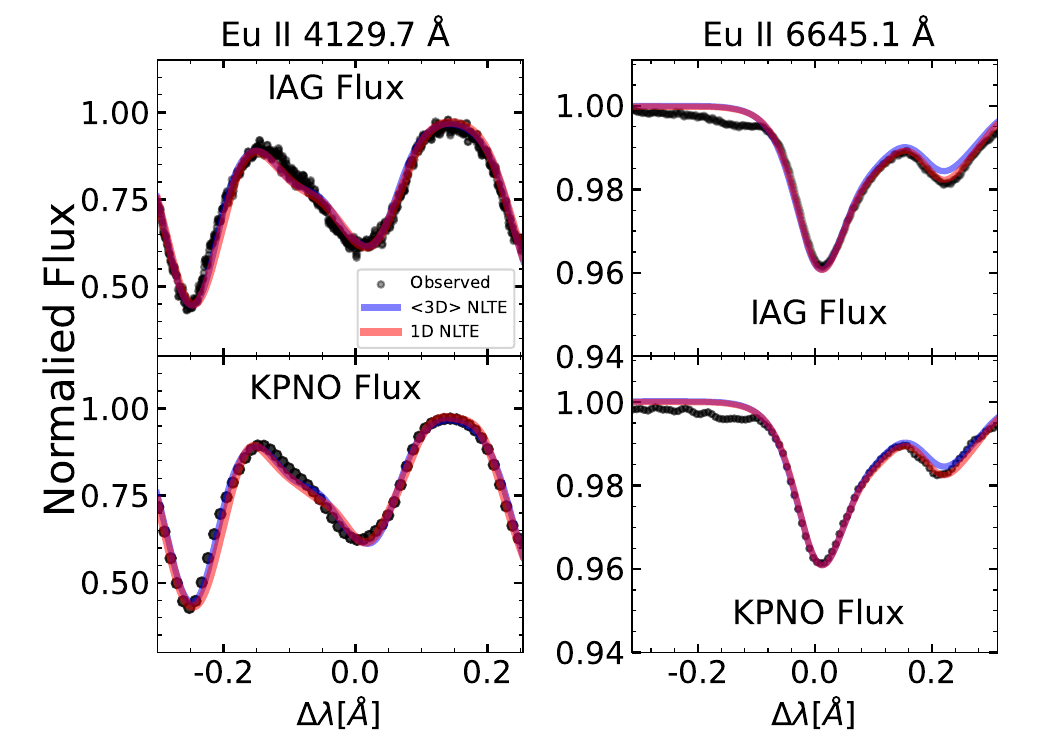}
   \caption{Best-fit synthetic spectra of sun for \GG{$\lambda$ 4129\AA\ (left panels)} and Eu II $\lambda$ 6645\AA\ (right panels) generated from TSFitPy. The red line and blue line represent the synthetic spectra based on 1D and $<$3D$>$ NLTE, respectively. The observed data are shown as black dots.}
    \label{fig: Eufit_sun_spectrum}
\end{figure} 

In the top panels of Fig.~\ref{fig: Eufit_sun_spectrum}, we show the observed Sun spectrum of the selected \GG{Eu II $\lambda$ 4129\AA\ line (left)} and Eu II $\lambda$ 6645\AA\ (right) line in black dots, taken from the Vacuum Vertical Telescope at the Institute f\"ur Astrophysik G\"ottingen
(IAG), with a resolving power of R = 700,000 \citep{2016Reiners_IAG}.
Here, we overplot the synthetic spectrum based on the 1D NLTE model with the best-fit \GG{A(Eu) = 0.61 $\pm 0.05
$ for $\lambda$ 4129\AA\ line} and A(Eu) = 0.55 $\pm 0.05$ for $\lambda$ 6645\AA\ line in red lines. 
The synthetic spectrum based on the $<$3D$>$ NLTE model with the best-fit \GG{A(Eu) = 0.64 $\pm 0.05
$ for $\lambda$ 4129\AA\ line} and A(Eu) = 0.58 $\pm 0.05$ for $\lambda$ 6645\AA\ line is shown as the blue lines. 
\GG{We accounted for the blending of the Eu II $\lambda$ 4129.7 \AA\ line by Ti I $\lambda$ 4129.66 \AA\ line and Sc I $\lambda$ $\lambda$4129.75 \citep{2000Mashonkina}.
For Eu II $\lambda$ 6645.70 \AA\ line, we accounted for the blending of 
 Si I $\lambda$ 6645.21 line.
For different stars in Sec.~\ref{sec:GCEresults}, the Ti, Sc and Si abundance was scaled to [Fe/H].}
The individual error components were calculated by adding the systematic uncertainty (including blends and uncertainty of $f$-values) and the fitting error in quadrature \citep{2024A&AStorm}.
In the bottom panels of Fig.~\ref{fig: Eufit_sun_spectrum}, we present the observed Sun spectrum in black dots, derived from the solar Kitt Peak National Observatory (KPNO) FTS atlas with R $\approx$ 400,000 \citep{1984Kurucz}.
The synthetic spectrum based on the 1D NLTE model with the best-fit \GG{A(Eu) = 0.62 $\pm 0.05
$ for $\lambda$ 4129\AA\ line} A(Eu) = 0.58$\pm 0.05$ for the $\lambda$ 6645\AA\ line are overplotted in red, while the synthetic spectrum based on the $<$3D$>$ NLTE model with the best-fit \GG{A(Eu) = 0.64 $\pm 0.05
$ for $\lambda$ 4129\AA\ line} and A(Eu) = 0.60 $\pm 0.05$ for $\lambda$ 6645\AA\ line are shown in blue.
This indicates that the Eu abundances derived from the 1D NLTE model are consistent across the two observed spectra, and similarly, the results from the $<$3D$>$ NLTE model are also consistent.
The best-fit values of 1D LTE and $<$3D$>$ LTE, are also provided in Table.~\ref{tab:sun Eu abundance}.

\begin{table*}
\caption{\label{tab:sun Eu abundance}Derived A(Eu) abundances based on 1D and $<$3D$>$ LTE/NLTE models for IAG and KPNO spectra.}
\centering
\begin{tabular}{ccccc}
 \hline \hline
Spectra   &   1D LTE & 1D NLTE & $<$3D$>$ LTE &$<$3D$>$ NLTE   \\ \hline
IAG flux (4129) &  0.54  &  0.61   & 0.59 & 0.64\\  
KPNO flux (4129)&  0.55  &  0.62   & 0.59 & 0.64\\  
IAG flux  (6645)&  0.54  &  0.55   & 0.60 & 0.58\\  
KPNO flux (6645)&  0.57  &  0.58   & 0.63 & 0.60\\  
\hline 
\end{tabular}\\
\end{table*}

Our results for 1D LTE and NLTE abundances are higher than those reported by \cite{2000Mashonkina}, who obtained LTE and NLTE abundances for $\lambda$ 4129\AA\ line are 0.49 and 0.53, while for the $\lambda$ 6645\AA\ line, these are 0.50 and 0.53, respectively.
The differences could be attributed to several factors, including the different loggf values used in our study (\GG{0.22 for $\lambda$ 4129\AA\ line} and 0.12 for $\lambda$ 6645\AA\ line from \citet{2001Lawler}; see Section \ref{sec: NLTE calculate}), compared to the value of \GG{0.174 for $\lambda$ 4129\AA\ } and 0.204 for $\lambda$ 6645\AA\ taken from \cite{1991Komarovskii} in \cite{2000Mashonkina}. 
Additionally, the Eu II lines are also influenced by hyperfine structure and isotopic shifts. 
We used the hyperfine structure data described in Section \ref{sec: Linelist} \G{for the Eu II $\lambda$ 6645\AA\ line, represented with 11 hyperfine and isotopic components from \citet{2024A&AStorm}}; whereas \cite{2000Mashonkina} used data from \cite{1976Biehl}.  
The Eu isotope ratio used in \cite{2000Mashonkina} was 55:45, whereas we used a ratio of 47.8:52.2 from \cite{2009Lodders} based on the Gaia-ESO line list \citep{2021Heiter}. 
Furthermore, we used the MULTI1D code to calculate the statistical equilibrium of Eu, while \cite{2000Mashonkina} used the NONLTE3 code.
In addition, due to these differences in atomic data and the statistical equilibrium code, \citet{2024A&AStorm} tested the main factors influencing the results and concluded that they are primarily due to the use of a more comprehensive atomic model.


Our 1D LTE/NLTE Eu abundances obtained from Eu II $\lambda$ 6645\AA\ line (0.57 $\pm 0.05$ and 0.58 $\pm 0.05$) are in good agreement with \citet{2024A&AStorm}, who reported values of 0.59$\pm 0.01_{\rm stat} \pm 0.06_{\rm syst}$ and 0.60 $\pm 0.01_{\rm stat} \pm 0.06_{\rm syst}$.
The 1D LTE/NLTE Eu abundances of IAG spectrum are slightly lower than their results, but still within their error bar. 
Although the loggf values \GG{and the 1D model atmosphere} used are the same, these differences could still arise from different radiative transfer codes, calculations of departure coefficients, as they use the MULTI3D@DISPATCH code \citep{2024Eitner}, while we use Turbospectrum and MULTI1D for NLTE calculations (see Sect.~\ref{sec: NLTE calculate}).
\GG{
The obtained discrepancies between these two codes are within 0.05 dex for the strong lines, e.g. Mn lines (Storm et al. in prep).
Similarly, \cite{2012BergemannFe} and \cite{2019BergemannMn} used the same model atom and got slightly different results using different codes.
}

The \G{difference} for the Eu II $\lambda$ 6645\AA\ line in abundance corrections is consistent with findings from \citet{2024A&AStorm}, showing an increasing between 1D LTE and 1D NLTE, while decreasing from $<$3D$>$ (full 3D) LTE to $<$3D$>$ (full 3D) NLTE.
However, our $<$3D$>$ results are higher than their full 3D results, which are 0.58 and 0.55. 
This may indicate that $<$3D$>$ cannot represent all of the 3D effects, but nevertheless provides a promising insight into its effects.

\subsection{NLTE effects on Eu II lines}
\begin{figure}
    \centering
    \includegraphics[scale=0.6]{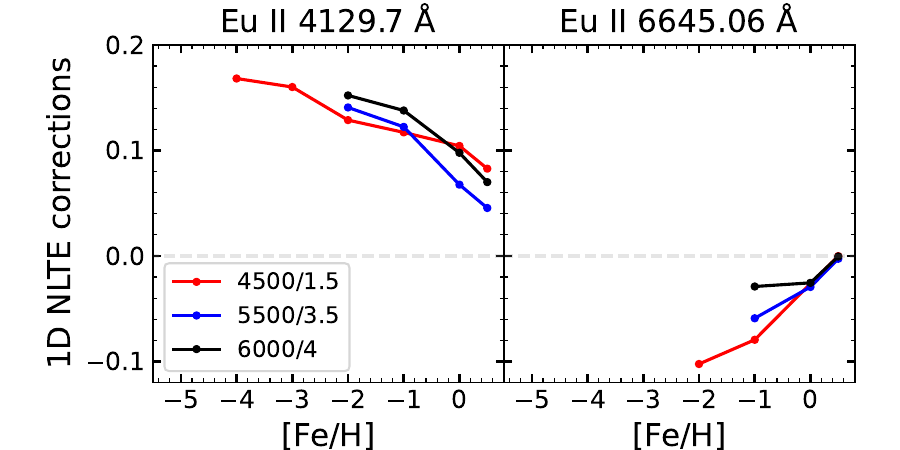}
    \caption{NLTE corrections for the Eu II \GG{$\lambda$ 4129\AA\ line (left panel)} and $\lambda$ 6645\AA\ line (right panel) are plotted against the metallicity [Fe/H] of the model atmospheres. 
    The red line represents the NLTE correction of red giant with parameters of $T_{\rm eff}$=4500 K/$\log{g}$=1.5 dex, the blue line represents the NLTE correction of subgiant with parameters of $T_{\rm eff}$=5500 K/$\log{g}$=3.5 dex, and the black line represents the NLTE correction of main-sequence star with parameters of $T_{\rm eff}$=6000 K/$\log{g}$=4.0 dex. \G{The NLTE calculations were performed assuming [Eu/Fe] = 0.}}
    \label{fig: NLTE correction}
\end{figure} 

Figure~\ref{fig: NLTE correction} shows the NLTE corrections for the Eu II \GG{$\lambda$ 4129\AA\ line (left panel)} and $\lambda$ 6645\AA\ lines (right panel) plotted against the metallicity [Fe/H] of the model atmospheres. 
The red line depicts the NLTE correction for red giants (RG) with parameters of $T_{\rm eff}$=4500/$\log{g}$=1.5, while the black line illustrates the NLTE correction for main-sequence stars (MS), with parameters of $T_{\rm eff}$=6000/$\log{g}$=4.0.
Additionally, the blue line represents the NLTE correction of subgiant (SG) with parameters of $T_{\rm eff}$5500/$\log{g}$3.5.
\GG{
The NLTE corrections for Eu II $\lambda$ 4129 \AA\ are positive, while negative or close to zero for Eu II $\lambda$ 6645 \AA\ .
For Eu II $\lambda$ 4129 \AA\ line, at values of [Fe/H] near $-2$ dex, the corrections for RG, SG, and MS are close to each other.}
For the Eu II $\lambda$ 6645 \AA\ line, the NLTE corrections for RG are slightly higher than for SG, and the corrections for SG are slightly higher than those for MS stars when [Fe/H] is less than 0 dex.
For the MS, the NLTE corrections do not exceed $-0.03$ dex at [Fe/H] = $-1$ dex. 
However, for RG, the NLTE corrections can reach $-0.1$ dex at [Fe/H]= $-2$ dex. 
It also indicates that the NLTE corrections increase with decreasing metallicity, which may have an impact on the GCE trend of [Eu/Fe] especially with the lower [Fe/H] \citep{2023Alexeeva,2024A&AStorm}.
\GG{Although the NLTE corrections for the Eu II $\lambda$ 4129 \AA\ and Eu II $\lambda$ 6645 \AA\ lines are entirely opposite, the final results after NLTE corrections match (as discussed in Section 4.4), further proving the necessity of NLTE corrections for determination of Eu abundance.}

\begin{figure}
    \centering
    \includegraphics[scale=0.6]{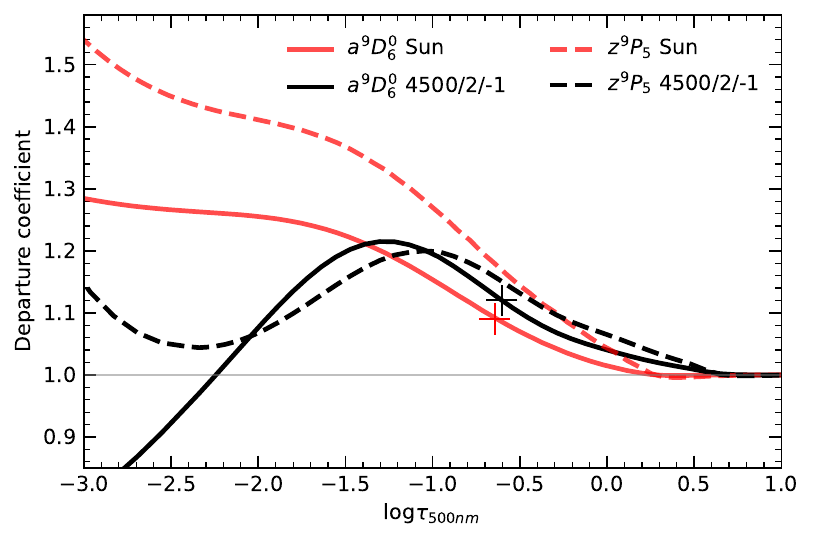}
    \caption{Departure coefficients for the Eu II $\lambda$ 6645 \AA\ line as a function of optical depth. The solar model atmosphere are depicted in red and $T_{\rm eff}=$4500 K/$\log{g}=$2.0 dex/[Fe/H]$=-1$ dex in black, with solid and dashed lines representing lower and upper levels, respectively.}
    \label{fig: Departure}
\end{figure}

We show the departure coefficients, $b_{i}$, for Eu II $\lambda$ 6645 \AA\ line as a function of optical depth in Fig.~\ref{fig: Departure}.
Here, the departure coefficient, $b_{i}$ =$\frac{n_{i}^{\text{NLTE}}}{n_{i}^{\text{LTE}}}$, is defined as the ratio of the NLTE population of atomic level $n_{i}^{\text{NLTE}}$ to the LTE population of atomic level, $n_{i}^{\text{LTE}}$.
The solid lines represent the lower levels, while the dashed lines represent the upper levels.
The departure coefficients for the solar model atmosphere are depicted in red, while the black one represents the model atmosphere with $T_{\rm eff} = 4500$ K, $\log{g} = 2.0$~dex, and [Fe/H] = $-1$ dex.
We chose this model atmosphere because its parameters are similar to those of most of our sample stars.
The formation height is shown as + \G{at log($\tau$) = 0 for line center.}
In the solar atmosphere, \G{the formation height of 6645 \AA\ line} corresponds to log($\tau_{500}$) = $-$0.64, where both states $z^{9}P_{5}^{0}$ and $a^{9}D_{6}$ tend to be overpopulated compared to LTE values, with $b_{j}$ being higher than $b_{i}$ at the same time. 
According to \cite{2014Bergemann}, this suggests that the line source function exceeds the Planck function ($S_{\nu}^{l}$/$B_{\nu}^{l}$ $\approx$ $b_{j}$/$b_{i}$ $> 1$).
As a result, the line becomes weaker under NLTE conditions relative to LTE.
This ultimately causes slightly positive NLTE corrections.
For the model atmosphere with $T_{\rm eff} = 4500$ K, $\log{g} = 2.0$~dex, and [Fe/H] = $-1$ dex, another effect plays a more important role. 
The departure coefficient of upper energy state is closer to that of lower energy state ($b_{i} \approx b_{j}$) at log($\tau_{500}$) = $-$0.6.
Thus, the enhanced line opacity compared to LTE (since $\kappa^{1} \approx b_{i} > $1), increases the number of absorbed photons. 
This enhanced absorption strengthens the line under NLTE conditions, leading to a negative NLTE correction.

\begin{figure}
	\centering
	\includegraphics[scale=0.5]{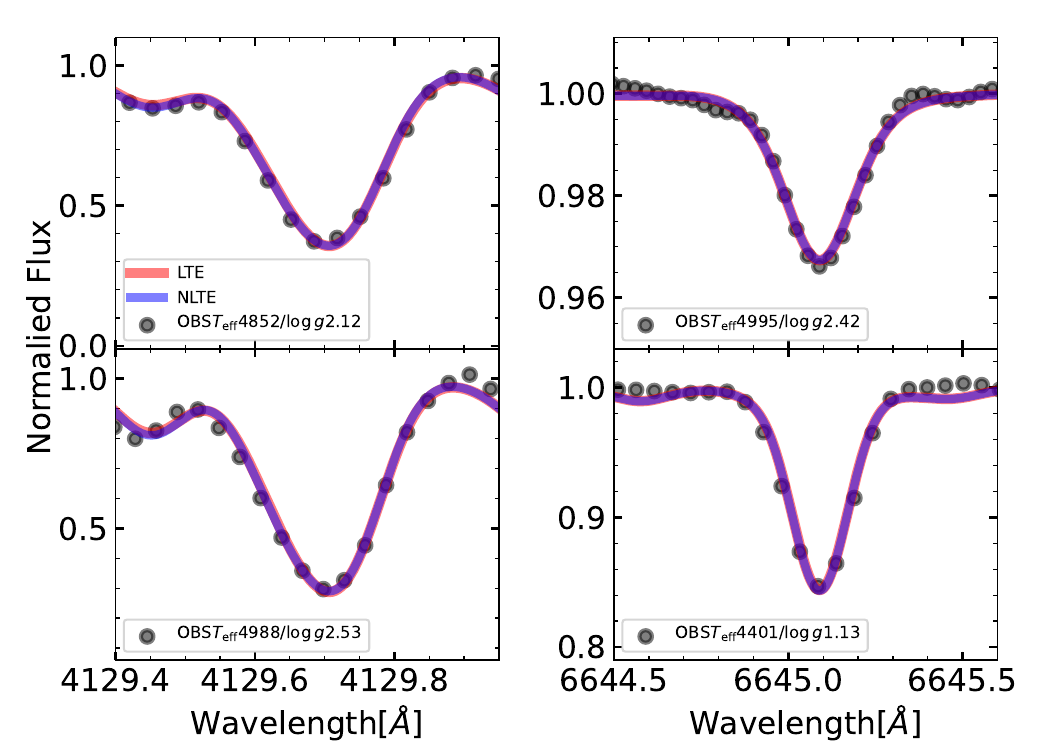}
    \caption{Examples of the fitting based on 1D LTE (red line) and NLTE (blue line) model. The black dots \G{in each panel} represent the observed spectrum with \G{four} different \G{stars}.}
    \label{fig: fitting_example}
\end{figure}

\begin{figure}
    \centering
    \includegraphics[scale=0.5]{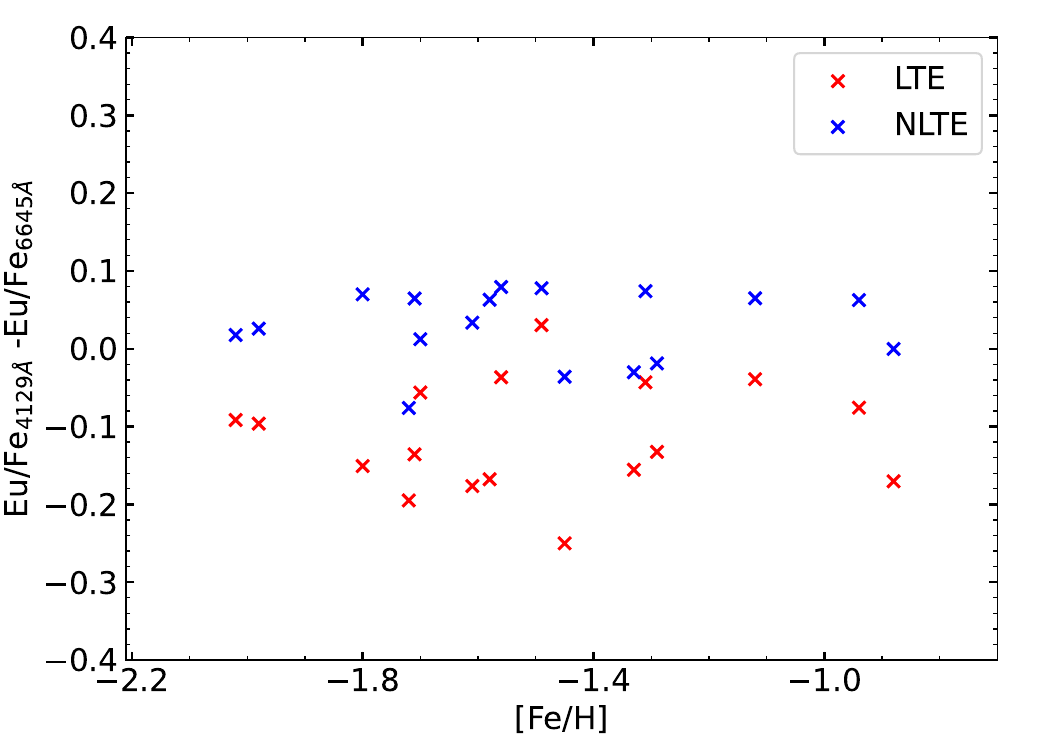}
    \caption{Comparison of Eu/Fe differences between the $\lambda$ 4129 \AA\ and $\lambda$ 6645 \AA\ lines under LTE (black) and NLTE (red) conditions. 
    }
    \label{fig: 6645vs4129}
\end{figure}

We present the fitting results of four stars as examples in Fig.~\ref{fig: fitting_example}.
For the Eu II 4129 \AA\ line, the observed spectrum is shown as black dots in the upper left panel, with parameters: $T_{\rm eff}$=4852 K, $\log{g}$=2.12 dex, and [Fe/H]=$-1.65$ dex; in the lower left panel, we have: $T_{\rm eff}$=4988 K, $\log{g}$=2.53 dex, [Fe/H]=$-1.43$ dex.
For the Eu II 6645 \AA\ line, we show the observed spectrum in black dots with parameters of $T_{\rm eff}$=4995 K, $\log{g}$=2.42 dex, and [Fe/H]=$-1.8$ dex in the upper right panel and $T_{\rm eff}$=4401 K, $\log{g}$=1.13 dex, and [Fe/H]=$-1.41$ dex in the lower right panel.
The red and blue lines represent the best-fit synthetic spectra based on 1D LTE and NLTE models, respectively.
We show the comparison of [Eu/Fe] differences between the $\lambda$ 4129 \AA\ and $\lambda$ 6645 \AA\ lines in Fig.~\ref{fig: 6645vs4129}.
The NLTE corrections result in more consistent values between the two lines, highlighting the importance of NLTE corrections for precise Eu abundance determinations.

\subsection{Comparison of the results with GCE model}\label{sec:GCEresults}
\begin{table*}
\caption{\label{tab:Eu abundance4129}Results of the Eu abundance based on Eu II 4129\AA\ line.}
\centering
\begin{tabular}{ccccccccccccc}
 \hline \hline
Spectra   &   $T_{\rm eff}$ & $\log{g}$ & [Fe/H] & $\xi_{t}$ & [Eu/Fe]$_{\rm 1D LTE}$ & [Eu/Fe]$_{\rm 1D NLTE}$  \\ \hline
RAVE J004733.4-450942  &  4925  &  2.18   & $-1.78$ & 1.5 & 0.42& 0.56\\  
RAVE J022308.0-312953  &  4783  &  1.93   & $-1.65$ & 1.9 & 0.56& 0.68\\  
RAVE J022658.3-074959  &  5125  &  2.84   & $-1.71$ & 1.2 & 0.43& 0.57\\
RAVE J031005.5-401003  &  5065  &  2.68   & $-1.35$ & 1.5 & 0.46& 0.58\\
RAVE J031535.8-094744  &  4774  &  2.06   & $-1.31$ & 1.6 & 0.58& 0.65\\
\hline 
\end{tabular}\\
\begin{center}
This table is available in its entirety in machine-readable form at the CDS.
\end{center}
\end{table*}

\begin{table*}
\caption{\label{tab:Eu abundance}Results of the Eu abundance based on Eu II 6645\AA\ line.}
\centering
\begin{tabular}{ccccccccccccc}
 \hline \hline
Spectra   &   $T_{\rm eff}$ & $\log{g}$ & [Fe/H] & $\xi_{t}$ & [Eu/Fe]$_{\rm 1D LTE}$ & [Eu/Fe]$_{\rm 1D NLTE}$  \\ \hline
TYC 8789-00425-1       &  4383  &  1.22   & $-1.56$ & 2.1 & 0.35& 0.30\\  
TYC 9161-00706-1       &  4992  &  2.02   & $-1.39$ & 1.6 & 0.49& 0.39\\  
RAVE J212558.5-265657  &  4795  &  2.04   & $-1.61$ & 1.9 & 0.52& 0.44\\
RAVE J204547.1-621403  &  4995  &  2.42   & $-1.8 $ & 1.5 & 0.75& 0.67\\
RAVE J104943.6-092646  &  4695  &  2.04   & $-1.13$ & 1.6 & 0.74& 0.72\\
\hline 
\end{tabular}\\
\begin{center}
This table is available in its entirety in machine-readable form at the CDS.
\end{center}
\end{table*}

\GG{The wavelength range for some of our stars does not include the Eu II 4129 \AA\  line, and in some cases, both the Eu II 4129 \AA\ } and Eu II 6645 \AA\ lines are blended or very weak.
Therefore, the results presented below are only for the spectra that were well fitted \GG{with a total of 164 stars having metallicities ranging from \GG{$-2.4$} to $-0.5$ dex}.
However, the $<$3D$>$ grid covers a smaller range of atmospheric parameters \citep{2013Magic1,2013Magic2} and, thus, only a small proportion of our sample of stars was fitted with these models. 
Therefore, our results are only based on the results of the 1D.
All the best fitting results are provide \GG{in Table~\ref{tab:Eu abundance4129} for Eu II 4129 \AA\ } and Table~\ref{tab:Eu abundance} for Eu II 6645 \AA\ .
There are five stars in our sample that are carbon-enhanced metal-poor \GG{with s-process element enhancement (CEMP-s)} stars \citep{2018Placco}, which do not follow the Galactic chemical evolution trend. 
We use * to mask them in Tables~\ref{tab:Eu abundance4129} and~\ref{tab:Eu abundance} .

\begin{figure}
    \centering
    \includegraphics[scale=0.5]{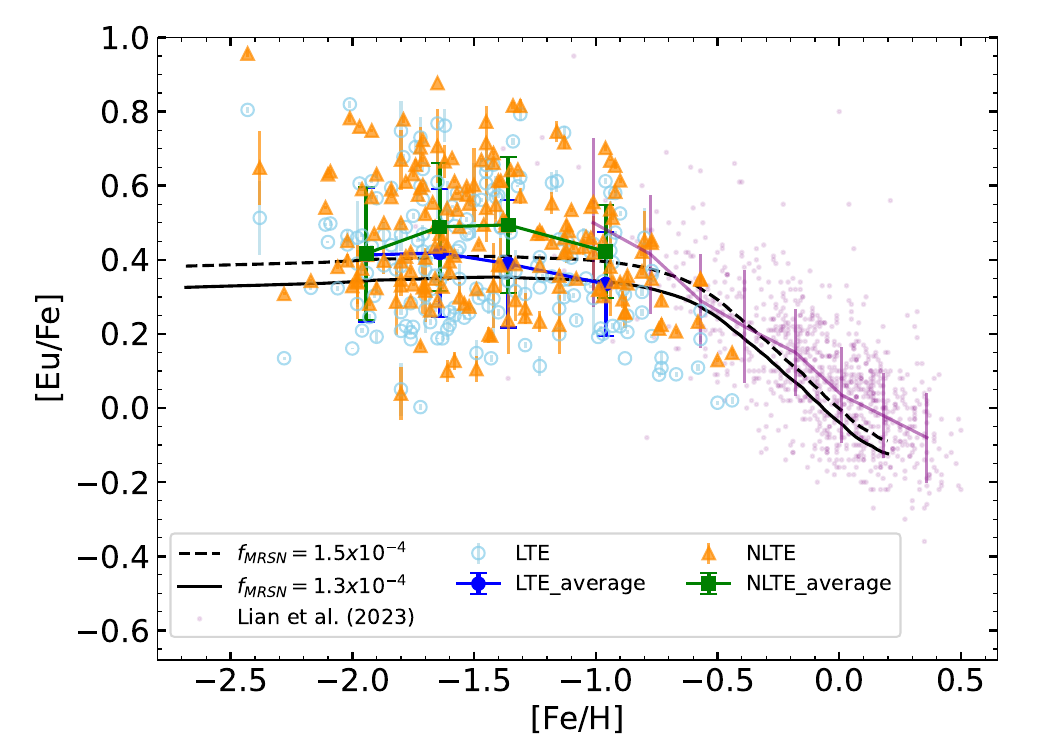}
    \caption{Trend of [Eu/Fe] based on the best-fit results of 1D NLTE and LTE. 
    \GG{The green squares and blue dots represent the averaged NLTE and LTE [Eu/Fe] ratios across the selected bins.
    The yellow triangles and cyan circles represent the NLTE and LTE [Eu/Fe] for each stars.
    The purple dots represent 1D LTE Eu abundance measurements of 1274 metal-rich stars from Gaia-ESO survey.
    }
    The dashed and solid lines represent the GCE models with different fraction of massive stars that end up with MRSN instead of CCSN. }
    \label{fig: Eu_trend}
\end{figure}

In Fig.~\ref{fig: Eu_trend}, we show the [Eu/Fe] based on 1D NLTE of our metal-poor sample as a function of [Fe/H] combined with 1D LTE Eu abundance measurements of 1274 metal-rich stars from Gaia-ESO survey \citep{2022Gilmore,2022Randich} with a high spectral signal-to-noise ratio (S/N$>50$; \citealt{2023Lian}). 
\GG{We divided the bins to ensure approximately an equal number of stars in each, with the green squares and blue dots representing the averaged NLTE and LTE [Eu/Fe] ratios across the selected bins.
The yellow triangles and cyan circles represent the NLTE and LTE [Eu/Fe] values for a total of 141 stars based on the $\lambda$ 4129 \AA\ line and 35 stars based on the $\lambda$ 6645 \AA\ line combined.} As shown in Fig.~\ref{fig: Eu_trend}, the trend of [Eu/Fe] is almost flat in metal-poor region and both the scatter of data points and the NLTE adjustments tend to decrease from low to high metallicity.
However, this phenomenon requires confirmation based on more data points in the future. 

For comparison, the tracks of two galactic chemical evolution (GCE) models generated by OMEGA+ \cite{cote2018} using the parameters of the basic model in \citet{2023Lian} are presented. In these models, to account for the rapid decrease of [Eu/Fe] in the metal-rich regime \G{which implies much shorter release timescale of Eu than Fe by SN-Ia}, the contribution of Eu on short timescale from magneto-rotating supernova (MRSN) has been taken into account, in addition to the neutron star merger (NSM) that releases Eu on much longer timescales \citep{cote2019}. For NSM, we assumed an ejecta mass of 2.5$\times10^{-2}{\rm M_{\odot}}$ with yields table from \citet{arnould2007} and occurrence rate of 2$\times10^{-5}$ for every solar mass of star formation. The delay time distribution of NSM is assumed to follow a power law in the form of $t^{-1}$ with minimum delay time of 10~Myr and maximum delay time of 10$^6$~Gyr. To include the contribution of MRSN, we replaced a tiny fraction (0.013\% and 0.015\%) of core-collapse supernova (CCSN) in the mass range of 13-25~${\rm M_{\odot}}$ by MRSN and used the yields table from \citet{nishimura2015}. This fraction is higher than that (0.01\%) adopted in \citet{2023Lian}, where the data in the low-metallicity regime is limited and the fraction of MRSN is not well constrained. 
\G{In this work we intend to test the impact of the NLTE corrections on the required parameters responsible for Eu production in GCE models. For simplicity, we focus on one of the key and poor constrained parameter, the fraction of MRSN in CCSN ($f_{\rm MRSN}$), which is positively correlated with the [Eu/Fe] abundance in old metal-poor stars. As expected, with higher [Eu/Fe] after NLTE correction, a GCE model with larger $f_{\rm MRSN}$ is needed to well fit the average [Eu/Fe] of our sample. We note that the required change in ($f_{\rm MRSN}$) is not substantial given the NLTE correction derived in this work.}


\section{Summary}\label{sec:Summary} 
As an r-process element, Eu provides insights into violent events such as neutron star mergers, helping us to improve our understanding of nucleosynthetic production sites. 
Due to the lack of observational data for r-process elements, especially for Eu, based on NLTE analysis in the metal-poor region, this work is aimed at investigating Eu abundances in metal-poor stars by applying NLTE and $<$3D$>$ corrections.

We performed a detailed analysis of the NLTE effects on Eu II 4129 \AA\ and Eu II 6645 \AA\ for a sample of metal-poor stars. 
For the Eu II 4129 \AA\ line, we found that the NLTE effects result in positive NLTE abundance correction in both 1D and $<$3D$>$ in the solar atmosphere.
For the Eu II 6645 \AA\ line, we found that the NLTE effects result in a slightly positive NLTE abundance correction in 1D and a negative one in $<$3D$>$ in the solar atmosphere.
\GG{Although the NLTE corrections for the Eu II $\lambda$ 4129 \AA\ and Eu II $\lambda$ 6645 \AA\ lines are entirely opposite, the discrepancy between the abundances derived from individual lines decreases after NLTE fitting, once again showcasing the importance of NLTE abundance determination.}
Finally, we show the trend of [Eu/Fe] as a function of [Fe/H] and compare it with the GCE models. 
It indicates that the NLTE correction does not \G{require significant change of the parameter for the Eu production} in GCE models. 
However, due to the limitations of observational data and models, confirming this phenomenon would require more data points, for instance, with 4MOST in the future.

\section{Data availability}\label{sec:Summary} 
Tables 3 and 4 are only available in electronic form at the CDS via anonymous ftp to cdsarc.u-strasbg.fr (130.79.128.5) or via http://cdsweb.u-strasbg.fr/cgi-bin/qcat?J/A+A/.

\begin{acknowledgements}
This work is supported by the Natural Science Foundation of China (Nos.\ 12288102,12125303,12090040/3,12103064,12403039), the National Key R\&D Program of China (grant Nos. 2021YFA1600403/1, 2021YFA1600400), and the Natural science Foundation of Yunnan Province (Nos. 202201BC070003, 202001AW070007), the International Centre of Supernovae, Yunnan Key Laboratory (No. 202302AN360001)
and the “Yunnan Revitalization Talent Support Program"-Science and Technology Champion Project (N0. 202305AB350003).
NS and MB acknowledge funding from the European Research Council (ERC) under the European Union’s Horizon 2020 research and innovation programme (Grant agreement No. 949173). 
MB is supported through the Lise Meitner grant from the Max Planck Society.
We sincerely thank Dr. Hongliang Yan for his valuable suggestions that improved the quality of this paper.
We acknowledge support by the Collaborative Research centre SFB 881 (projects A5, A10), Heidelberg University, of the Deutsche Forschungsgemeinschaft (DFG, German Research Foundation).  
\end{acknowledgements}

\bibliographystyle{aa} 
\bibliography{paper} 
\end{document}